\documentclass[11pt]{article}
\usepackage[a4paper,hmarginratio=1:1,vmarginratio=2:3,totalwidth=15.8cm,
totalheight=21.5cm]{geometry}
\usepackage{bm,epstopdf,epsfig,amsmath,amssymb,amsfonts,
latexsym,comment,cancel,verbatim}
\usepackage[rflt]{floatflt}
\usepackage[font=md,captionskip=8pt]{subfig}
\usepackage{amsmath}
\usepackage{amssymb}
\usepackage{amsfonts}
\usepackage{epsfig}
\usepackage{graphicx}
\long
\def\symbolfootnote[#1]#2{\begingroup
\def\thefootnote{\fnsymbol{footnote}}
\footnote[#1]{#2}\endgroup} 
\providecommand{\U}[1]{\protect\rule{.1in}{.1in}}

\def\beq{\begin{equation}}
\def\eeq{\end{equation}}
\def\bea{\begin{eqnarray}}
\def\eea{\end{eqnarray}}
\def\bq{\begin{quote}}
\def\eq{\end{quote}}

\def\bq{\begin{quote}}
\def\eq{\end{quote}}

\renewcommand{\baselinestretch}{1.25}
\begin{document}

\begin{flushright}
OUTP-0924P\\
CERN-PH-TH-2009-209
\end{flushright}

\thispagestyle{empty}
\vspace{2.5cm}
\begin{center}
{\Large \textbf{Testing SUSY}}
\vspace{1.cm}

\textbf{S. Cassel$^{\,\,a}$, 
D.~M. Ghilencea$^{\,\,b,\!}$\symbolfootnote[1]{on leave from 
Theoretical Physics Department, IFIN-HH Bucharest MG-6, Romania.}, 
G. G. Ross$^{\,\,a,\,b,}$\symbolfootnote[2]{e-mail addresses:
s.cassel1@physics.ox.ac.uk, dumitru.ghilencea@cern.ch,
g.ross1@physics.ox.ac.uk}} \\[0pt]

\vspace{0.5cm} {$^a\, $ Rudolf\, Peierls\, Centre for Theoretical Physics,
\,University\, of\, Oxford,\\[0pt]
1 Keble Road, Oxford OX1 3NP, United Kingdom.}\\[6pt]
{\ $^b\, $ Department of Physics, CERN - Theory Division, CH-1211 Geneva 23,
Switzerland.}\\[6pt]
\end{center}

\vspace{1cm}

\noindent
\begin{abstract}
If SUSY provides a solution to the hierarchy problem then supersymmetric
states should not be too heavy. This requirement is quantified by a fine tuning measure that provides a quantitative test of
SUSY as a solution to the hierarchy problem. The measure is useful in
correlating the impact of the various experimental measurements relevant to
the search for supersymmetry and also in identifying the most sensitive
measurements for testing SUSY. In this paper we apply the measure to the
CMSSM, computing it to two-loop order and taking account of current
experimental limits and the constraint on dark matter abundance. Using this
we determine the present limits on the CMSSM parameter space and identify
the  measurements at the LHC that are most significant in covering the remaining
parameter space. Without imposing the LEP Higgs mass bound we show
that the smallest fine tuning (1:13) consistent with a relic density within the WMAP bound corresponds to a Higgs mass of 114$\pm$2 GeV. Fine tuning rises rapidly for heavier Higgs.
\end{abstract}

\newpage
\setcounter{page}{1}
\section{Introduction}

As the Large Hadron Collider starts operation, the search for
supersymmetry reaches the interesting stage at which candidate models for
physics beyond the Standard Model (SM) may be definitively tested. In order
to do this it is necessary to quantify the viable range of the SUSY\
parameters, and in particular limit the mass scale of the supersymmetric
spectrum. The bounds on the masses of supersymmetric states come most
sensitively\footnote{Light
 supersymmetric states are also needed for accurate gauge coupling
unification. However the dependence on the mass is only logarithmic so gauge
unification does not provide the most restrictive bound on the SUSY\
spectrum.} from the requirement that supersymmetry solves the hierarchy
problem, i.e. it ensures that the electroweak breaking scale is consistent
with radiative corrections without undue fine tuning. Fine tuning 
\cite{Ellis:1986yg}-\cite{Athron:2007qr} 
is a measure of the probability of unnatural cancellations between soft
SUSY\ breaking terms in the determination of the electroweak breaking scale
after including quantum corrections and experimental constraints. The LEP\
constraints have already placed SUSY\ under some pressure as the fine tuning
is found to be quite large. The reason for this is largely due to the LEP
bound on the mass of the lightest Higgs mass, $m_{h},$ because satisfying it
requires large radiative corrections that depend logarithmically on the mass
of the top squarks and this in turn requires large stop masses and hence
large supersymmetry breaking soft terms\footnote{Unless the Higgs has
 non-standard decays \cite{Chang:2008cw}.}. However, even in the
simplest implementations of SUSY, there is still a
significant area of parameter space to explore and it is of importance 
to provide a quantitative measure of what needs to be done to fully test SUSY.

In this paper we shall use the fine tuning measure, $\Delta$, to quantify
fine tuning. 
The measure determines the cancellation needed between the
independent terms contributing to the electroweak symmetry breaking vacuum
expectation value and provides an intuitively reasonable way of quantifying
fine tuning. Of course it is necessary to supplement this by imposing an
upper bound for $\Delta$ beyond which fine tuning is unacceptable and this
is a subjective judgement. For the most part we will just display the range
of fine tuning found without imposing such a cut off, except when assessing
whether the LHC will be able fully to test the model. As we shall discuss
the fine tuning measure allows one to quantify the significance of
individual measurements in testing SUSY over the full parameter space and so
is useful in correlating the various measurements relevant to SUSY\
searches. It also allows us to determine the most important measurements for
testing SUSY at the LHC.

In order to illustrate the usefulness of the fine tuning analysis we will
study in detail the 
Constrained Minimal Supersymmetric Model (CMSSM)
which is the minimal supersymmetric extension of the SM with a restricted set of
SUSY\ breaking soft terms. We later comment on the effect of relaxing
this assumption and on other effects beyond CMSSM.
 We compute $\Delta$ to two-loop order, using the
results of 
\cite{Espinosa:2000df}-\cite{Carena:1995bx} 
and present a scan of the full CMSSM parameter space compatible with current
experimental and theoretical constraints. In particular we require:

\begin{itemize}
\item radiative generation of electroweak symmetry breaking.

\item no colour/charge breaking vacua.

\item consistency with current experimental bounds on superpartner masses,
electroweak precision data, $b\rightarrow s\gamma$, $b\rightarrow \mu \mu$ 
and anomalous magnetic moment constraints.

\item a radiatively corrected SM-like Higgs mass in agreement with the
current LEPII bound on $m_{h}$.

\item consistency with the thermal relic density constraint. 
\end{itemize}

We start with the form of the scalar potential responsible for electroweak
breaking. With the standard two-higgs doublet notation, it is given by
\medskip
\begin{align}
V &
=m_{1}^{2}\,\,|H_{1}|^{2}+m_{2}^{2}\,\,|H_{2}|^{2}-(m_{3}^{2}\,\,H_{1}\cdot
H_{2}+h.c.)  \notag \\[6pt]
& ~+~\frac{1}{2}\,\lambda_{1}\,|H_{1}|^{4}+\frac{1}{2}\,\lambda_{2}
\,|H_{2}|^{4}+\lambda_{3}\,|H_{1}|^{2}\,|H_{2}|^{2}\,+\lambda_{4}\,|H_{1}
\cdot H_{2}|^{2}  \notag \\[5pt]
& ~+~\bigg[\,\frac{1}{2}\,\lambda_{5}\,\,(H_{1}\cdot H_{2})^{2}+\lambda
_{6}\,\,|H_{1}|^{2}\,(H_{1}\cdot
H_{2})+\lambda_{7}\,\,|H_{2}|^{2}\,(H_{1}\cdot H_{2})+h.c.\bigg]
\label{2hdm}
\end{align}

\medskip\noindent with the assumption that at the UV scale 
$m_1^2=m_2^2=m_0^2+\mu_0^2$, and $m_3^2=B_0^{} \, \mu_0^{}$.

The couplings $\lambda_{j}$ and the soft masses receive one- and two-loop
corrections that for the MSSM are found in \cite{Martin:1993zk,Carena:1995bx}
. We shall use these results to evaluate the overall amount of fine-tuning
of the electroweak scale. To this purpose, it is convenient to introduce the
notation 
\begin{align}
m^{2} &
=m_{1}^{2}\,\cos^{2}\beta+m_{2}^{2}\,\sin^{2}\beta-m_{3}^{2}\,\sin2\beta \\%
[6pt]
\lambda & =\frac{\lambda_{1}^{{}}}{2}\,\cos^{4}\beta+\frac{\lambda_{2}^{{}}}{
2}\,\sin^{4}\beta+\frac{(\lambda_{3}^{{}}+\lambda_{4}^{{}}+\lambda_{5})}{4}
\,\sin^{2}2\beta+\sin2\beta\left( \lambda_{6}^{{}}\cos^{2}\beta
+\lambda_{7}^{{}}\sin^{2}\beta\right)  \label{ml}
\end{align}

\medskip\noindent Minimisation of $V$ gives \medskip 
\begin{equation}
v^{2}=-m^{2}/\lambda,\qquad2\lambda\frac{\partial m^{2}}{\partial\beta}=m^{2}
\frac{\partial\lambda}{\partial\beta}  \label{min}
\end{equation}

\medskip\noindent which fixes $v$ and $\beta$ as functions of the MSSM bare
parameters.

The fine-tuning measure, $\Delta$, is defined by \cite{Ellis:1986yg}
\medskip
\begin{equation}
\Delta\equiv\max\big\vert\Delta_{p}\big\vert_{p=\{
\mu_{0}^{2},m_{0}^{2},m_{1/2}^{2},A_{0}^{2},B_{0}^{2}\}},\qquad\Delta_{p}
\equiv\frac {\partial\ln v^{2}}{\partial\ln p}  \label{ft}
\end{equation}
where all $p$ are input parameters at the UV scale, in the standard MSSM
notation\footnote{We 
have also studied the contribution to $\Delta$ coming from the
uncertainty in the top Yukawa coupling and the strong coupling. Using the
modified definition~\cite{Ciafaloni:1996zh} of $\Delta$ appropriate to
measured parameters we find them to be sub-dominant.}.

We compute the fine tuning at two-loop order including the dominant third
generation supersymmetric threshold effects to the scalar potential. The
analysis is done in two stages. First a scan is performed over all of
parameter space using a slightly simplified two-loop calculation developed
to run quickly. Then the analysis is redone using the (slower) \texttt{
SOFTSUSY~3.0.10} package~\cite{Allanach:2001kg} that includes all the
effects mentioned above for a set of points that has low fine tuning. It was
found important to work at two-loop order as $\Delta$ changes significantly
between one and two-loops. 
Note that \texttt{SOFTSUSY} uses the measured values of the gauge couplings. The unification scale is determined by the point the $SU(2)$ and $U(1)$ couplings unify (of $O(10^{16}$~GeV). 
Full details of this procedure and an analysis of
the results will appear in a separate publication~\cite{fulldetails}. The
results are shown in Figure~1. In this the Higgs mass 
and $\Delta$ are computed by \texttt{SOFTSUSY}. 
It agrees to within $0.1$\,GeV
with that determined by \texttt{SuSpect}~\cite{Djouadi:2002ze} but can
differ by $\pm2$\,GeV~\cite{Allanach:2004rh}
from that determined by \texttt{FeynHiggs}~\cite{Hahn:2009zz}. Given this
uncertainty, which comes from the higher order terms in the perturbative
expansion, the LEP\ bound\footnote{Strictly
 this has been derived for the SM only. However it also applies to
the CMSSM for viable parameter choices.}~\cite{higgsboundLEP} should be
interpreted as $m_{h}>114.4\pm2$\,GeV. 
In what follows we will usually quote fine tuning computed for the central
value. The relic density constraint is tested using \texttt{MicrOMEGAs 2.2}~\cite{Belanger:2006is}.
\begin{figure}[ht]
\def\baselinestretch{1.}
\begin{center}
\subfloat[Relic density unrestricted]{\includegraphics[width=7.1cm]{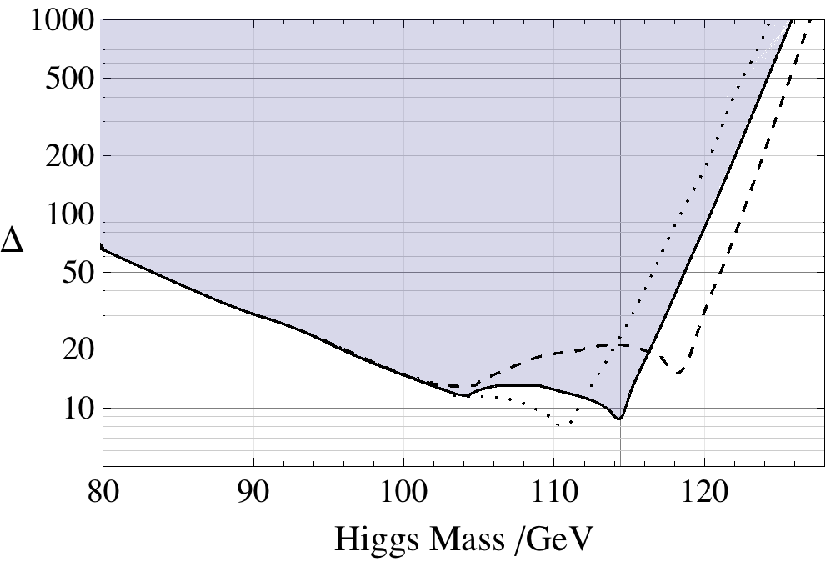}} 
\hspace{7mm} 
\subfloat[Points satisfying WMAP
bound]{\includegraphics[width=7.1cm]{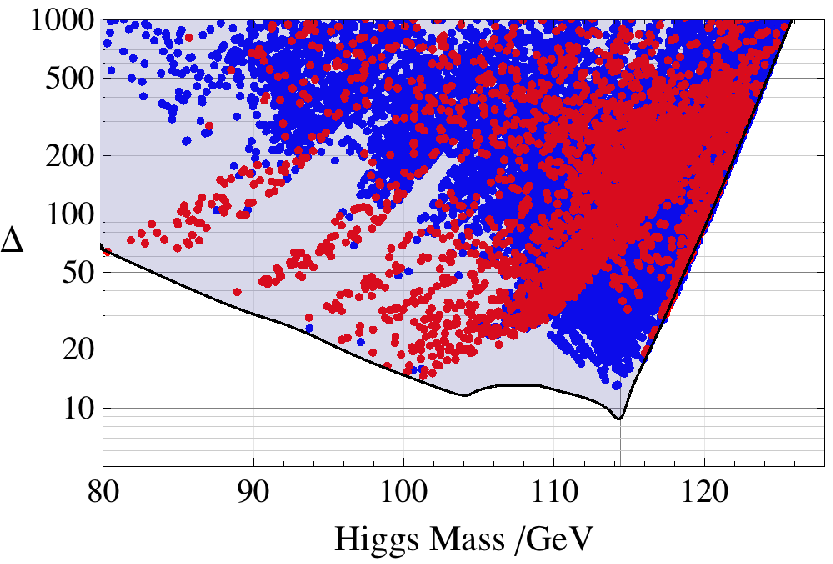}}
\end{center}
\caption{{\protect\small Two-loop fine tuning vs Higgs mass for the scan $2
\leq \tan \protect\beta \leq 55$. The solid line is the minimum fine tuning
with $\left( \protect\alpha_s^{} , M_t^{} \right) =$ (0.1176, 173.1~GeV).
The dashed and dotted lines in (a) have $\left( \protect\alpha_s^{} , M_t^{}
\right) =$ (0.1156, 174.4~GeV) and (0.1196, 171.8~GeV) respectively to
account for the 1$\protect\sigma$ experimental errors. In (b), the red
(lighter) points saturate the non-baryonic relic density within the 
1$\protect\sigma$ WMAP bounds, $\Omega h^{2}=0.1099\pm0.0062$. The blue
(darker) points have a thermal relic density, $\Omega h^{2} < 0.1037$. }}
\label{higgsp}
\end{figure}

The fine tuning distribution shows an interesting structure with a 
minimum close to the present LEP\ bound. This structure is largely generated
by the $\lambda $ dependence of $v$ in eq(\ref{min}). At tree level the MSSM
value for $\lambda $ is anomalously small, $\lambda
=(g_{1}^{2}+g_{2}^{2})\cos ^{2}\left( 2\beta \right) /8,$ and this is the
main reason why fine tuning is large at tree level. Radiative corrections
can increase $\lambda $, reducing the fine tuning.
The structure shown in Figure~1 is driven by this effect. For $m_{h}$ less
than the LEP\ bound the fine tuning is dominated by the $\mu $ contribution
and rises for \textit{decreasing} $m_{h}$. This is because the region corresponds to smaller $\tan \beta$
where the radiatively corrected  $\lambda $ is smaller. Since $v^2=-m^2/\lambda$, we see from eq(5) that smaller $\lambda$ leads to an increase
in fine tuning  (for an extended discussion see \cite{fulldetails}).
For $m_{h}$ greater than the LEP\ bound the fine tuning is
dominated by the $m_{0}$ contribution because in this region it is necessary
to have large positive radiative corrections to the Higgs mass and this
requires large stop masses and in turn large $m_{0}$, greater than the focus
point~\cite{Feng:2000bp} can control. Since the Higgs mass depends
logarithmically on the stop masses, in this region $m_{0}^{2}$ is roughly
proportional to $e^{m_{h}^{2}}.$ The fine tuning measure in this region is
in turn approximately proportional to $m_{0}^{2}$ and hence grows
exponentially with the Higgs mass. The two-loop corrections increases this
growth slightly, 
explaining the sharp rise seen in Figure~1 at large $m_{h}$. The position of
the dip in fine tuning comes from the lower bounds on the SUSY\
masses; as they increase the dip moves to higher masses and $\Delta $ at the
minimum increases. If it rises beyond an unacceptable level one may conclude
that the CMSSM solution to the hierarchy problem has been fully tested and
found to fail. At present we are far from this point with regions of SUSY\
parameter space having fine tuning less than 1 part in 8.8. 

Not yet included in the analysis is the constraint coming from the dark
matter abundance. In Figure~1(b) 
we show that there are points in parameter space that populate the regions
with small fine tuning and have a dark matter density consistent with
providing all or some of the dark matter. 
Note that this favours the part of the dip around the LEP\ Higgs
mass bound ($c.f.$\cite{Chankowski:1998za},\cite{Ellis:2007by})
 even though this bound has \textit{not} been included in the
analysis!  Taking into account the theoretical uncertainty in
determining 
$m_{h}$ we conclude from that the most likely mass for the Higgs consistent
with the observed dark matter abundance is $m_{h}=114\pm 2$ GeV
corresponding to a fine tuning of $1:13$. For a saturation of the WMAP bound within $1\sigma$, one finds $m_h = 116\pm2$ GeV corresponding to a fine tuning of 1:19. For $m_{h}=121$ GeV this
rises to 
$1:100$ fine tuning and at $m_{h}=126$ GeV to $1:1000$ fine tuning.
\begin{figure}[t!h!]
\def\baselinestretch{1.1}
\center
\subfloat[Fine Tuning vs $\tan \beta$]{\includegraphics[width=6.7cm] {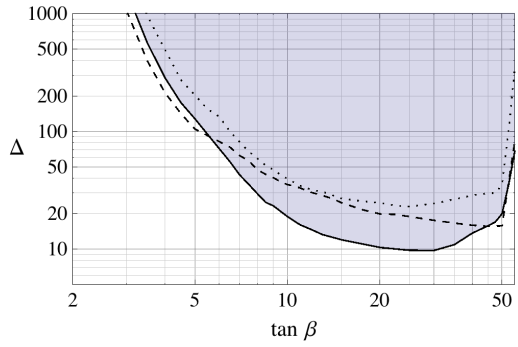}}
\hspace{7mm}
\subfloat[Fine Tuning vs
$A_0^{}$]{\includegraphics[width=6.7cm]{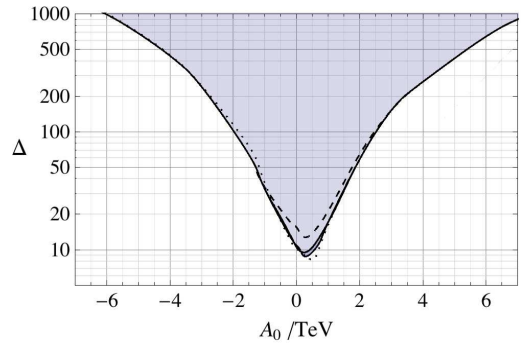}}
\hspace{3mm} 
\medskip
\subfloat[Fine Tuning vs $m_0^{}$]{\includegraphics[width=6.7cm] {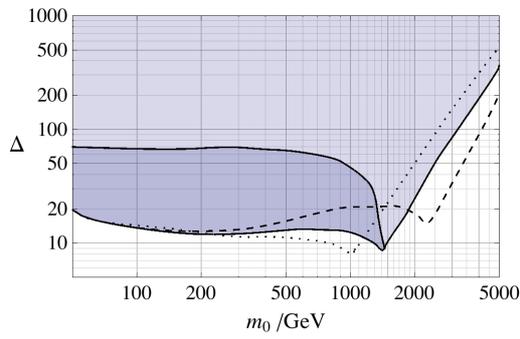}} 
\hspace{7mm}
\subfloat[Fine Tuning vs
$m_{1/2}^{}$]{\includegraphics[width=6.5cm]{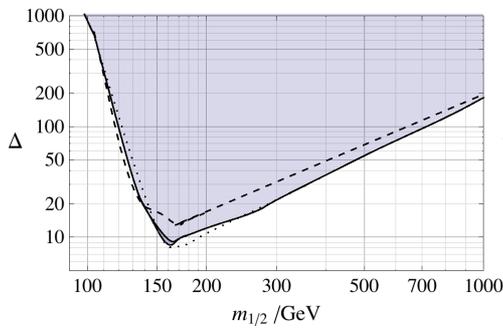}}
\medskip
\subfloat[Fine Tuning vs $\mu$]{\includegraphics[width=6.7cm]{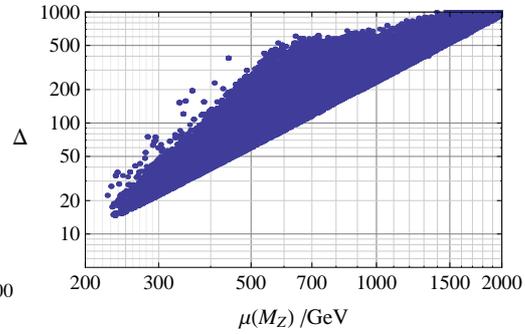}}
\caption{{\protect\small Dependence of minimum fine tuning on SUSY
parameters ($\protect\mu >0$, relic density unrestricted). The solid, dashed and dotted lines are as explained in Fig~\protect\ref{higgsp}. In (b)-(d), the
darker shaded regions are eliminated when $m_{h}^{{}}>114.4$~GeV is
applied for the case with the central $\left( \protect\alpha 
_{s}^{{}},M_{t}^{{}}\right) $ values. In (a) and (e), $m_h > 114.4$\,GeV is applied, and the points in (e) are only for the central $\left( \protect\alpha _{s}^{{}},M_{t}^{{}}\right) $ values.}}
\end{figure}

The fine tuning measure can readily be used to establish the remaining
allowed range of the SUSY\ parameters by plotting the contribution to fine
tuning of the various components. This is shown in Figure~2. 
From these graphs one may see that, once one chooses an upper limit for the
fine tuning measure $\Delta ,$ the allowed range for the parameters is
defined. For the case that we define $\Delta =100$ as the 
upper limit beyond which we consider SUSY\ has failed to 
solve the hierarchy problem, we obtain the bounds:
\begin{equation}
\begin{tabular}{rlcrcl}
$m_{h}$ & $<~121$~\mbox{GeV} & \hspace{5mm} & $5.5~<$ & $\tan \beta $ & $<~55
$ \\[3pt]
$\mu $ & $<~680$~\mbox{GeV} &  & 120~\mbox{GeV}$~<$ & $m_{1/2}$ & $<~720$~
\mbox{GeV} \\[3pt]
$m_{0}$ & $<~3.2$~\mbox{TeV} &  & $-2$~\mbox{TeV}$~<$ & $A_{0}$ & $<~2.5$~
\mbox{TeV}
\end{tabular}
\end{equation}

It is clear that there is still a wide range of parameters that needs to be
explored when testing the CMSSM. Will the LHC be able to cover the whole
range? To answer this note that one must be able to exclude the upper limits
of the mass parameters appearing in Table 1. Of course the state
that affects fine tuning most is the Higgs scalar and one may see from
Figure~1 that establishing the bound $m_{h}>121$\,GeV will imply that 
$\Delta>100.$ However the least fine tuned region corresponds to the lightest
Higgs consistent with the LEP\ \ bound and this is the region where the LHC
searches rely on the $h\rightarrow\gamma\gamma$ channel which has a small
cross section and will require some $30\,fb^{-1}$ at $\sqrt{s}=14$\,TeV to
explore. Given this it is of interest to consider to what extent the direct
SUSY\ searches will probe the low fine tuned regions. 

For $\mu $ the $\Delta =100$ upper bound corresponds to a Higgsino mass of $%
O(0.5$\thinspace TeV) while for $m_{1/2}$ the bound corresponds to a gluino
mass of $O(1.5$\thinspace TeV) and a Wino or neutralino mass of $O(300$
\thinspace GeV). 
On the other hand, due to the focus point behaviour, the bound on $m_{0}$ is
very weak corresponding to squark and slepton masses in the multi TeV
region. Thus we expect that in testing SUSY the most significant processes
at the LHC will be those looking for gluinos, winos and neutralinos. In
Table~1 we give the upper limits on the mass of these states corresponding
to $\Delta =100.$ For comparison we show the limit on the stop and sbottom
masses, much larger due to the weak limit on $m_{0}.$ The remaining states
of the MSSM\ spectrum are dominated by the $m_{0}^{{}}$ soft mass term and
are bounded by $O$(3~to~4.4~TeV). \ All these upper mass limits scale
approximately as $\sqrt{\Delta _{\mbox{\tiny min}}^{{}}}$, so may be adapted
depending on how much fine tuning the reader is willing to accept.
\smallskip
\begin{table}[ht]
\begin{center}
\begin{tabular}{|c||c|c|c|c||c|c||c|c||c|c|}
\hline
$\tilde{g}$ & $\chi_{1}^{0}$ & $\chi_{2}^{0}$ & $\chi_{3}^{0}$ & $%
\chi_{4}^{0}$ & $\chi_{1}^{\pm}$ & $\chi_{2}^{\pm}$ & $\tilde{t}_{1}^{}$ & $%
\tilde{t}_{2}^{}$ & $\tilde{b}_{1}^{}$ & $\tilde{b}_{2}^{}$ \\ \hline\hline
1720 & 305 & 550 & 660 & 665 & 550 & 670 & 2080 & 2660 & 2660 & 3140 \\ 
\hline
\end{tabular}
\end{center}
\caption{\small 
CMSSM upper mass limits on superpartners (in GeV),
such that $\Delta<100$ remains possible.}
\label{partlimit}
\end{table}

Studies of SUSY\ at the LHC~$\cite{Baer:2009dn}$ have shown that in the LHC
experiments have a sensitivity to gluinos of mass $1.9$\thinspace TeV for 
$\sqrt{s}=10$\thinspace TeV, $2.4$\thinspace TeV for $\sqrt{s}=14$\thinspace
TeV and luminosity $10fb^{-1}.$ Of relevance to the first LHC\ run the limit
is $600$\thinspace GeV for $\sqrt{s}=10$\thinspace TeV and luminosity 
$100\,pb^{-1}.$ These correspond to probing up to $\Delta =120,180$ and $14$
respectively. 
As we have discussed charginos and neutralinos can be quite light, but their
signal events are difficult for LHC to extract from the background, owing in
part to a decreasing $M_{\widetilde{W}}-M_{\widetilde{Z}}$ mass gap as 
$\left\vert \mu \right\vert $
decreases~\cite{Barbieri:1991vk,Baer:2004qq}. An Atlas study
\cite{Vandelli:2007zza}  of the trilepton signal from
chargino-neutralino production found that $30fb^{-1}$ luminosity at
14 TeV is needed for a
 3$\sigma$ discovery significance for $M_2<300$~GeV and $\mu<250$~GeV \cite{CMS}.

Finally we return to the intriguing fact that minimum fine tuning plus
correct dark matter abundance corresponds to a Higgs mass just
 above the LEP\ bound. As we
have noted above this point is fixed by the current bounds on the SUSY\
spectrum and not by the current Higgs mass bound which is not included when
doing the scans leading to Figure~1. One may interpret the SUSY\ parameters
corresponding to this point as being the most likely given our present
knowledge and so it is of interest to compute the SUSY\ spectrum for this
parameter choice as a benchmark for future searches. This is presented in
Table~2 where it may be seen that it is somewhat non-standard with very
heavy squarks and sleptons and lighter gauginos. This has some similarities
to the SPS2 scenario \cite{BA}. 

\bigskip

\begin{table}[ht]
\begin{center}
\begin{tabular}{|c|c||c|c||c|c||c|c|}
\hline
$h^{0}$ & 114.5 & $\tilde{\chi}_1^0$ & 79 & $\tilde{b}_{1}^{}$ & 1147 & $%
\tilde{u}_{L}^{}$ & 1444 \\[1pt] 
$H^{0}$ & 1264 & $\tilde{\chi}_2^0$ & 142 & $\tilde{b}_{2}^{}$ & 1369 & $%
\tilde{u}_{R}^{}$ & 1446 \\[3pt] 
$H^\pm$ & 1267 & $\tilde{\chi}_3^0$ & 255 & $\tilde{\tau}_{1}^{}$ & 1328 & $%
\tilde{d}_{L}^{}$ & 1448 \\[2pt] 
$A^0$ & 1264 & $\tilde{\chi}_4^0$ & 280 & $\tilde{\tau}_{2}^{}$ & 1368 & $%
\tilde{d}_{R}^{}$ & 1446 \\[2pt] 
$\tilde{g}$ & 549 & $\tilde{\chi}_1^\pm$ & 142 & $\tilde{\mu}_L^{}$ & 1406 & 
$\tilde{s}_{L}^{}$ & 1448 \\[2pt] 
$\tilde{\nu}_{\tau}^{}$ & 1366 & $\tilde{\chi}_2^\pm$ & 280 & $\tilde{\mu}%
_R^{}$ & 1406 & $\tilde{s}_{R}^{}$ & 1446 \\[2pt] 
$\tilde{\nu}_{\mu}^{}$ & 1404 & $\tilde{t}_{1}^{}$ & 873 & $\tilde{e}_L^{}$
& 1406 & $\tilde{c}_{L}^{}$ & 1444 \\[2pt] 
$\tilde{\nu}_{e}^{}$ & 1404 & $\tilde{t}_{2}^{}$ & 1158 & $\tilde{e}_R^{}$ & 
1406 & $\tilde{c}_{R}^{}$ & 1446 \\ \hline
\end{tabular}
\end{center}
\par
\caption{A favoured
 CMSSM spectrum ($\Delta = 14.7$). Masses are given in $GeV$.}
\label{favspec2}
\end{table}

The results we have presented apply to the case of the CMSSM. It is
natural to ask how these results are modified when one considers
additional effects beyond CMSSM or relaxes some of its constraints.
For example, one could consider  different UV boundary
conditions for scalar masses, non-universal  gaugino masses or
 UV  threshold corrections to the gauge couplings. Let us
discuss briefly how  these can change our findings.
Universal scalar masses lead to a minimum of fine tuning through a focus point \cite{Feng:2000bp} and the minimum fine tuning found here corresponds to this focus point. Thus we expect fine tuning will increase if the universal scalar mass assumption is relaxed. For the case of non-universal gaugino masses, it is known that 
a reduction of the amount of fine tuning can be obtained 
\cite{Kane:1998im} below that of the CMSSM case. 
However this condition is not always sufficient to reduce $\Delta$ on its
own, and depends on other parameters values such as $\mu$ or $m_0$, 
which usually dominate the fine tuning \cite{fulldetails}. 
For a recent discussion of the reduction in fine tuning that can result from non-unversal gaugino masses  see  \cite{Horton:2009ed}.
Regarding UV scale threshold corrections to gauge couplings, 
if one imposes gauge coupling unification, they can affect the predictions for the low energy values of the couplings
 ($\alpha_s(M_Z)$, ...) \cite{Ghilencea:1997mu,Mayr:1993kn,Mayr:1995rx}
and hence, {\it c.f.} Figure 1, the fine tuning. However the bottom-up \texttt{SOFTSUSY} approach used here uses the measured values for the gauge couplings  and hence does not need to know the UV scale threshold corrections. The fine tuning due to the running of the soft SUSY breaking parameters is only midly sensitive to the UV scale threshold corrections.

In summary we have made a detailed study of the possibility of testing the
CMSSM as a solution to the hierarchy problem using a fine
tuning measure. This provides a way of quantifying and correlating the
effects of the numerous experimental measurements sensitive to SUSY and for
determining the region of parameter space that needs to be explored when
testing the model. Using this analysis\footnote{It 
is hoped to provide in the near future a simple computer package to allow
the analysis to be easily updated. In the meantime the authors will be happy
to provide the updated information on request.} new measurements providing
stricter bounds on the CMSSM spectrum will further limit the viable
parameter space. Similar analyses could (and should) be applied to more
general SUSY\ extensions of the SM and indeed to all proposed models
claiming to solve the hierarchy problem.

\section*{Acknowledgements}

The research was partially supported by the EU ITN grant UNILHC 237920
(``Unification in the LHC era'').
SC is supported by the UK Science and Technology Facilities Council
(PPA/S/S/2006/04503). D.G. thanks the CPhT-\'Ecole Polytechnique for kind
hospitality and financial support through the ERC Advanced Grant -
226371 (``MassTeV''), during the final stages of this work.
He also thanks S. Pokorski for interesting discussions on this topic.

\end{document}